\documentclass[fleqn,10pt]{wlscirep}

\usepackage{ifthen}
\usepackage{ifpdf}
\usepackage{color}

\ifpdf
\usepackage{graphicx}
\usepackage{epstopdf}
\else
\usepackage{graphicx}
\usepackage{epsfig}
\fi
\graphicspath{{./Figs_SBM/}{./}}

\usepackage{latexsym}
\usepackage{amsmath}
\usepackage{amssymb}
\usepackage{bm}
\usepackage{wasysym}

\usepackage{mathptmx}
\DeclareSymbolFont{epsilon}{OML}{ntxmi}{m}{it}
\DeclareMathSymbol{\epsilon}{\mathord}{epsilon}{"0F}

\usepackage{hyperref}

\newcommand{\const}{\mbox{const}}

\newcommand{\mass}{\mathsf{m}}

\newcommand{\eexp}[1]{\mathrm{e}^{#1}}

\newcommand{\braket}[1]{ \left\langle #1 \right\rangle}

\newcommand{\be}[1]{\begin{eqnarray}\ifthenelse{#1=-1}{\nonumber}{\ifthenelse{#1=0}{}{\label{e#1}}}}
\newcommand{\beq}{\begin{eqnarray}}
\newcommand{\eeq}{\end{eqnarray}} 

\newcommand{\hide}[1]{}

\newcommand{\Eq}[1]{\textcolor{blue}{{equation}\!~(\ref{#1})}} 
\newcommand{\Fig}[1]{\textcolor{blue}{Fig.}\!\!~\ref{#1}}
\newcommand{\sect}[1]{{\bf #1.-- }}

\newcommand{\rmrk}[1]{{#1}}     
\newcommand{\hrefl}[1]    {\href{#1}{[link]}}
\newcommand{\hidea}[1]{}    

\title{Superfluidity and Chaos {in low dimensional circuits}}

\author[1,*]{Geva Arwas}
\author[2,*]{Amichay Vardi}
\author[1,*]{Doron Cohen}

\affil[1]{Department of Physics, Ben-Gurion University of the Negev, Beer-Sheva, Israel}
\affil[2]{Department of Chemistry, Ben-Gurion University of the Negev, Beer-Sheva, Israel}

\begin{abstract}
The hallmark of superfluidity is the appearance of ``vortex states" carrying a quantized metastable circulating current. 
Considering a unidirectional flow of particles in a ring, at first it appears
that any amount of scattering will randomize the velocity, as in the Drude model, and eventually 
the ergodic steady state will be characterized by a vanishingly small fluctuating current. 
However, Landau and followers have shown that this is not always the case.  
If elementary excitations (e.g. phonons)  
have higher velocity than that of the flow, 
simple kinematic considerations imply metastability 
of the vortex state:  
the energy of the motion cannot dissipate into phonons.
On the other hand if this {\em Landau criterion} 
is violated the circulating current can decay.
Below we show that the standard Landau and Bogoliubov superfluidity criteria fail in low-dimensional circuits. Proper determination of the superfluidity regime-diagram must account for the crucial role of chaos, an ingredient missing from the conventional stability analysis. Accordingly, we find novel types of superfluidity, associated with irregular or chaotic or breathing vortex states. 
\end{abstract}

\begin{document}
\flushbottom
\maketitle
\thispagestyle{empty}

\section*{Introduction}

\sect{Metastability}
\rmrk{The essence of superfluidity is the possibility to witness 
a macroscopically-large persistent circulating flow of particles.}   
The Landau criterion \cite{Landau,Feynman,Hakim,Leboeuf,e1} 
\rmrk{for such metastability is determined by checking for accessible elementary excitations (phonons, rotons)} connecting the initial state to the quasi-continuum of other states with the same energy. \rmrk{The absence of an energetically-accessible pathway to all the other states with the same energy is termed {\em energetic-stability}.}
However, in low dimensional rings superfluidity can persist even if energetic-stability is lost \cite{dys1,dys2,dys5,dys6,Altman}. \rmrk{Namely, even in the absence of a ``potential-barrier", one may expect circumstances such that {\em dynamical stability} of the superfluid motion is maintained} (see also \cite{dys3,dys4}).  
\rmrk{But even if the superfluid motion become unstable and chaotic, it does not mean that free traffic between all the regions of the energy-shell is allowed:} Kolmogorov-Arnold-Moser (KAM) surfaces might effectively block the passage between different phase-space regions; More generally, remnants of integrable structures may allow only a slow percolation-like penetration process, so-called Arnold diffusion. These observations suggest a new paradigm of metastability as discussed below.

\sect{Low dimensional circuits}
The recent experimental realization of confining potentials with toroidal shapes and tunable weak links \cite{e8,e13,e15,n5,boshir2} has opened a new arena of studying superfluidity in low dimensional rings with Bosons.
Integrable (non-chaotic) 1D rings with Bosons have been extensively studied as models for superfluid circuits \cite{Udea,Brand1,Hekk}. 
\rmrk{However, dimensionality can be further reduced by considering a discrete $M$~site ring
\cite{rot1,rot2,rot3,brand3,ref9}. Discrete site systems have been realized experimentally in a wide range of setups, from double-well traps \cite{n1,Jeff} to large ($M{\gg}1$) arrays \cite{n2,dys6,BHH1,BHH2,e9}.
With regard to a ring geometry, of particular interest is the recent experimental realization of $M$~site rings using the `painted potential’ or related techniques \cite{ref9,e17}.}
For such ring the continuous rotational invariance is broken, hence chaos becomes a predominant feature. 
The system is formally equivalent to a set of $M$~coupled oscillators. \rmrk{Due to number-of-particles conservation, it is possible to separate one coordinate (and its conjugate) so as to have ${d=M{-}1}$ coordinates and their conjugates (say $d$ occupation differences and $d$ relative phases). Accordingly the effective number of degrees of freedom becomes~${d}$.}

Intensive studies have focused on the integrable Bosonic Josephson Junction (${M{=}2}$ hence ${d{=}1}$). In few-site Bose-Hubbard systems with $M{>}2$, such as the ${M{=}3}$ trimer\cite{ref12,trimer2,trimer3,trimer4,trimer6,trimer15,trimer7,trimer12,trimer13,trimer18,trimer19,trimer20}, an utterly new perspective is essential due to the emergence of chaos. So far, the implications of chaos on superfluidity have not been illuminated, neither for the trimer ring \cite{sfs} nor for $M{>}3$ rings. We would like to highlight a crucial 
difference between the two cases:  
For $M{=}3$ ring the $d{=}2$ dimensional KAM tori in phase-space
can divide the $2d{-}1$ energy-shell into separate regions, 
while for $M{>}3$ (say $M{=}4$, hence $d{=}3$) it is not the case (a 3D surface cannot divide a 5D energy-shell into disjoint territories). Consequently for ${M>3}$ all phase-space regions are interconnected via ``Arnold diffusion".     

\sect{Beyond the traditional view}
The traditional criterion for superfluidity associates the metastability of the current 
with the existence of a stationary {\em stable fixed point} of the Hamiltonian flow, 
that supports a coherent vortex state.
Accordingly a Bogoliubov de Gennes (BdG) linear-stability-analysis is assumed valid   
for determination of the regimes where superfluidity is anticipated. 
Low dimensional circuits have phase-space with both chaotic and quasi-regular motion. 
Consequently the traditional BdG paradigm is challenged: 
{\bf (i)}~Dynamical instability of a vortex state does not necessarily mean that
superfluidity is diminished, because its collapse may be topologically arrested by KAM structures;   
{\bf (ii)}~Linear BdG stability of a vortex state does not always imply actual stability, 
because Arnold diffusion can provide detour paths out of seemingly elliptical regions;   
{\bf (iii)}~Due to quantum fluctuation, i.e. finite uncertainty width of a vortex-state,
stability is required within a Plank cell around the fixed-point.  

The result of these three observations is a novel superfluidity regime-diagram, 
quite distinct from the one that would be obtained using standard criteria. 
%
Considering that ${d=M{-}1}$, and that Arnold diffusion can only take place when $d{>}2$, there should be a dramatic difference between trimers ($M{=}3$) and larger rings ($M{>}3$): 
For the trimer, item~(i) implies that superfluidity can persist 
even if the motion becomes {\em chaotic}; 
For larger rings, item~(ii) implies that BdG (linear) dynamical stability 
is not a sufficient condition; while item~(iii) implies 
that global analysis of phase-space topography is essential.  
In the extreme limit of ${M\rightarrow\infty}$ one should 
remember that the dynamics become integrable due to rotational symmetry.   
%
%
%
Below we demonstrate how the above ideas affect the regime diagram 
of a few site ring. For the trimer (${M=3}$) superfluidity manifests itself 
beyond the regime of dynamical stability, 
while for ${M>3}$ we find a much more intricate situation.

\section*{The Model and the superfluidity regime diagram} 

Consider $N$ mass~$\mass$ Bosons in an $M$~site ring that 
has a radius~$R$. If the ring is rotated with frequency~$\Omega$,   
one may transform into the rotating frame, where the potential is time-independent. 
In this frame we have Coriolis force, \rmrk{which is formally like having 
a magnetic flux $\Phi= (2\pi\mass R^2/\hbar)\Omega$} through the ring. 
Accordingly the system is described by the Bose-Hubbard Hamiltonian \cite{BHH1,BHH2} 
\be{1}
\mathcal{H} = \sum_{j=1}^{M} \left[
\frac{U}{2} a_{j}^{\dag} a_{j}^{\dag} a_{j} a_{j} 
- \frac{K}{2} \left(\eexp{i(\Phi/M)} a_{j{+}1}^{\dag} a_{j} + \text{h.c.} \right)
\right]~.
\eeq
Here $j \, \text{mod}(M)$ labels the sites of the ring, $a_{i}$ and $a_{i}^{\dagger}$ 
are destruction and creation operators. $K$ is the hopping frequency, 
and $U$ is the on-site interaction. 
\rmrk{The Peierls phase factor includes the so-called Sagnac phase $\Phi$.
Optionally it can be realized by introducing a spatially-adiabatic variation 
of the atomic magnetic dipole orientation \cite{e16}, or gauge field as in \cite{synt1}, see \cite{synt2}.} 
Without loss of generality ${\Phi\in[0,\pi]}$, and ${K>0}$, and ${U>0}$. 
Negative $U$ is the same as positive $U$ with a flipped energy landscape (${\mathcal{H} \mapsto -\mathcal{H}}$).
Negative $K$ is the same as positive $K$ with ${\Phi\mapsto\Phi+\pi}$ for odd~$M$.
Negative $\Phi$ is related to positive $\Phi$ by time reversal.  
The Hamiltonian $\mathcal{H}$ commutes with the total particle number $\mathcal{N}=\sum_{i}a^{\dagger}_{i}a_{i}$, 
hence the {\em operator} $\mathcal{N}$ is a constant of motion.
In a semi-classical context one defines phase-space 
action-angle coordinates as follows: 
\be{2} \nonumber
&& a_j \ \ = \ \ \sqrt{\bm{n}_j} \eexp{i\bm{\varphi}_j} 
\\ \nonumber
&& z \ = \ (\bm{\varphi}_1,\cdots,\bm{\varphi}_M,\bm{n}_1,\cdots,\bm{n}_M)
\eeq 
The Hamiltonian (\ref{e1}) is then expressed as $\mathcal{H}=H(z)$, 
and the dynamics is generated by $\dot{z} = \mathbb{J} \bm{\partial} H$. 
The notation $\partial_{\nu}$ stands for derivative with respect 
to $z_{\nu}$, and  
\beq \nonumber
\mathbb{J} \ \ = \ \  
\left( 
\scriptscriptstyle{
\begin{array}{cccc}
0 & \mathbb{I} \\
-\mathbb{I} & 0 \\
\end{array}
}
\right) 
\eeq
In the classical equation of motion, after standard 
rescaling of the variables, there are only 
two dimensionless parameters: one is the dimensionless 
interaction
\beq
u \ \ = \ \ {NU}/{K}
\eeq
and the other is the phase $\Phi$. Note that the re-scaling 
of the canonical variables implies that $\bm{n}$ is replaced 
by $\bm{n}/N$. Hence upon quantization $\bm{\varphi}$ and $\bm{n}$ 
are conjugate with dimensionless Plank constant $\hbar=1/N$.

\sect{The superfluidity regime-diagram} 
In \Fig{fg1} we plot the numerically determined ${(\Phi,u)}$ regime diagram 
for the superfluidity of rings with ${M=3,4,5}$ sites.
Image colors depict the current $I=\braket{(-\partial \mathcal{H}/\partial\Phi)}$ 
for the eigenstate that carries maximal current.
The solid and dashed lines indicate the energetic and the dynamical stability borders,  
as determined from the BdG analysis (see below).
The regime diagrams do not agree with the traditional analysis: 
For the $M{=}3$ ring superfluidity persists beyond 
the border of dynamical stability, while for $M{>}3$ 
the dynamical stability condition is not sufficient.

In the absence of interaction ($U{=}0$), time reversal is broken 
by having a non-zero $\Phi$, and an eigenstate can carry 
a non-zero ``persistent current". 
If we add some weak disorder~$W$ (random on-site potential)  
this current becomes smaller, and it diminishes in the limit $\Phi\rightarrow0$. 
In contrast superfluidity features a {\em macroscopically} large metastable current  
that is achieved due to having a non-zero interaction~($U{\ne}0$). 
Superfluidity is feasible if a middle vortex state (see below), 
or some irregular variant of it, maintains stability; 
otherwise it would mix with all the other eigenstates 
that reside in the same energy-shell, resulting in a micro-canonically small current.
Accordingly, the stability of the current should be verified with respect 
to an added disorder~$W$. In the model under study superfluidity 
is indeed maintained as long as ${W<U}$, 
and accordingly a {\em finite} strength of interaction~$U$ is required.


\section*{Metastability of vortex states}

\begin{figure}[h!]

\begin{center}

\includegraphics[width=1\hsize]{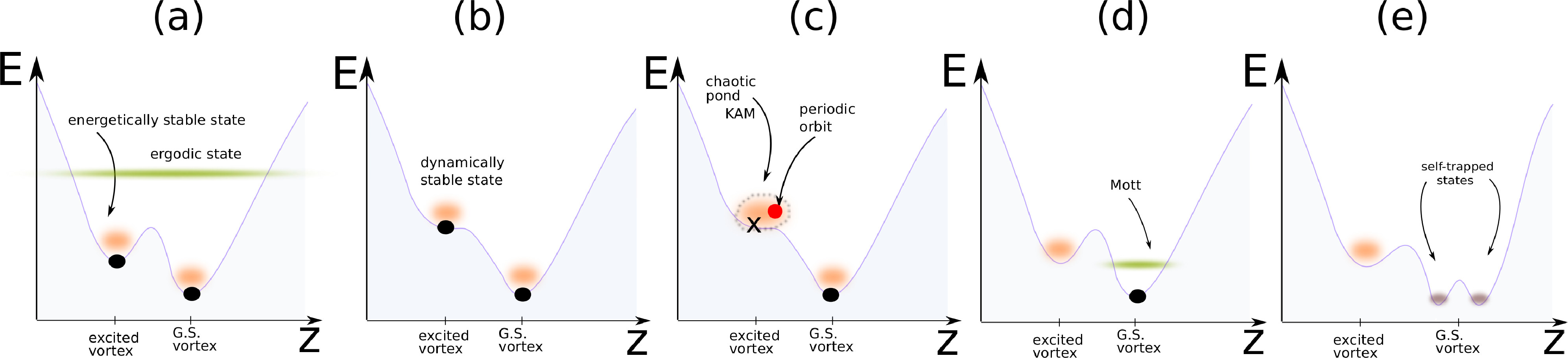}

%

\caption{\label{fg00} 
{\bf Different types of locked trajectories.} 
Locked trajectories form a cloud that can support a vortex state, 
or some irregular variant of a vortex state. 
This is merely a pedagogical cartoon (for a serious numerical illustration 
see the Poincare sections of \Fig{fg2}).     
Trajectories that are locked in a energetically-stable island, say a local minimum as in~(a), 
can support a regular vortex state; 
likewise with trajectories that are locked in a dynamically stable island of an elliptic fix-point as in~(b).
Trajectories that are locked in a chaotic pond, surrounding an hyperbolic fixed-point, 
can support a chaotic vortex state as in~(c). 
Another option in~(c) is to have trajectories that are locked to a nearby periodic orbit, 
supporting a breathing vortex state. 
For completeness, referring to the ground-state (GS) vortex,  
we illustrate in panels~(d) and~(e) the Mott transition, 
and the formation of self-trapped states (solitons).       
}

\end{center}
\end{figure}

The stationary orbitals of a single particle in a clean ring 
are the momentum states with wavenumber $k=(2\pi/M)m$, 
where $m$ is an integer modulo~$M$.
Coherent vortex states have $N$ particles {\em condensed}  
into the same momentum orbital. 
From a semiclassical perspective, a coherent-state
is represented by a Gaussian-like phase-space distribution.
Such state is stable if it is supported by a region 
where the classical motion is ``locked".   
\Fig{fg00} is a cartoon that summarizes various possibilities.
We first consider the traditional possibility of having 
a regular vortex-state that is supported by a stable fixed-point.  
In later sections we discuss the possibility of having  
irregular vortex-states that are supported by chaotic regions.

\sect{Stability analysis} 
A stable stationary fixed point of the classical Hamiltonian 
can support a coherent eigenstate. 
A stationary fixed point is the solution 
of the equation ${\bm{\partial} H=0}$.
A fixed-point is {\em energetically} stable if it 
resides at a local minimum or maximum of $H(z)$. 
\rmrk{Accordingly energetic-stability is determined 
by the eigenvalues of the 
Hessian matrix ${\mathcal{A}_{\nu,\mu} = \partial_{\nu} \partial_{\mu} H}$.}
In the vicinity of a fixed point the \rmrk{linearized} equation 
of motion takes the form  $\dot{z} = \mathbb{J} \mathcal{A} z$, 
\rmrk{Accordingly the nature of the dynamics   
is determined by the eigenvalues $\lambda_q$ of the 
monodromy matrix $\mathbb{J} \mathcal{A}$}. 
The latter are the roots of the characteristic 
equation ${\det(\lambda - \mathbb{J} \mathcal{A})=0}$.
Upon quantization $\omega_q= i\lambda_q$ are identified 
as the energies of the Bogoliubov excitations.
\rmrk{If the $\omega_q$ are real, the linearized motion is ``elliptic" around the fixed-point. But if some of the $\omega_q$ acquire an imaginary part, 
the motion becomes unstable (``hyperbolic"), 
meaning that trajectories depart exponentially 
from the fixed point. In the latter case a chaotic motion with a non-zero Lyapunov exponent is implied.}

\rmrk{In the present context there is a cyclic degree of freedom 
that corresponds to~$\mathcal{N}$.} Accordingly there is a pair  
of zero eigenvalues that should be excluded. Effectively we deal 
with $2d{\times}2d$ matrix where $d=M{-}1$).
It is also useful to notice that at the energetic-stability border, 
where the fixed-point becomes a saddle, 
we have ${\det(\mathcal{A})=\det(\mathbb{J}\mathcal{A})=0}$.
Vortex states correspond to the trivial fixed-points of the 
Hamiltonian,  located along the 
symmetry axis  $n_1=\cdots=n_M=N/M$. The $m$th vortex state,  
with $\varphi_i-\varphi_{i{-}1} = (2\pi/M)m$,  
corresponds to condensation in the $m$th momentum orbital.  
The vortex states have a macroscopically large current:
\beq
I_m \ \ = \ \ \braket{-\frac{\partial \mathcal{H}}{\partial \Phi}}_m
\ \ = \ \ \frac{N}{M}K \, \sin\left(\frac{1}{M}(2\pi m{-}\Phi)\right)
\eeq
Our linear-stability analysis of the vortex states \rmrk{is analogous to that of} \cite{dys1}, and will be presented in the following sections, \rmrk{with additional details in the Methods section,} leading to the solid and dashed lines of \Fig{fg1}.
%

\sect{Stability with respect to added disorder}
Before we proceed we would like to make a comment regarding the implications of having weak disorder.
The essence of superfluidity is the meta-stability of the current. 
This makes it distinguished from having merely a {\em persistent current}. 
Consider a clean ring with non-interacting particles and no disorder. 
We can prepare there a vortex state with macroscopically large current. 
However, any small {\em disorder} will randomize the velocity (as in Drude model) and the current will diminish. 
Adding interaction $U$ between the particles change the picture dramatically.
Now the energy landscape $E=H(z)$ has a non-trivial topography. It may have (say) 
a valley with a local minimum, that can support a metastable state.
If we add disorder it merely deforms the valley, while the fixed-point remains stable. 
Disregarding dimensionless prefactors this is true as long as ${W<U}$.
It follows that finite~$U$ is essential in order to have physical metastability. 
We would like to emphasize that the same conclusion holds for any ``energy landscape". 
Adding small disorder will affect only ``degenerated" regions, say the separatrix region, 
but will not affect the overall phase-space structure. This is called ``structural stability".    
To be on the safe-side, we have verified that the numerically determined 
regime diagrams are not affected by adding weak disorder.

\sect{Quantum fluctuations}
As shown in \Fig{fg1} the observed superfludity regimes are not in accordance 
with the traditional BdG analysis. Irrespective of~$M$, 
as we go higher in~$u$, superfluidity is diminished even in the energetically-stable region.
This is conspicuous for low $N$, as in \Fig{fg1}c where we have $\sim2$ particles per site, 
and can be explained as the consequence of having a finite uncertainty width.
Namely, as $u$ is increased the radius of the stability island (if exists) 
becomes smaller, until eventually it cannot support a stable vortex state. 
Equivalently, as~$N$ becomes smaller the uncertainty width of a  
vortex state becomes larger, until it 'spills' out of the stability island. 
This type of reasoning resembles the semiclassical view of the Mott transition (see below).   
Taking a closer look at the regime diagrams one observes that the above ``quantum fluctuations" 
perspective is not enough in order to explain the observed differences. We therefore turn to provide 
a more detailed phase-space picture.

\sect{The ground state} 
The lowest energy fixed-point ($m=0$ vortex state) 
is stable for any positive~$u$. However it is located in an island 
which is surrounded by a chaotic sea. As $u$ is increased the island's size 
decreases until at ${u>N^2/M}$, it becomes smaller than a single Planck cell.
At this point it can no longer accommodate a vortex state and one observes 
a quantum Mott transition. See \Fig{fg00}d for pedagogical illustration. 
Other studies of rotating ring lattices \cite{rot1,rot2,rot3,brand3,Hekk,ref9} 
have addressed additional quantum issues, such as the appearance of ``cat states":
In the trimer with $\Phi=\pi$ the ground state might be a macroscopic superposition 
of the degenerate vortex states $m{=}0$ and $m{=}1$.

\sect{Solitons}
While our main focus is on the stability of vortex states we briefly discuss 
other fixed points that were of interest in past work.
The trimer model without rotation (${\Phi=0}$) 
has been the subject of intense study 
\cite{ref12,trimer2,trimer3,trimer4,trimer6,trimer15,trimer7,
trimer12,trimer13,trimer18,trimer19,trimer20}. 
In particular it has been noted that the Hamiltonian $H(z)$ 
has additional non trivial fixed-points away from the symmetry axis.
The simplest of which are self-trapped  ``bright solitons" 
obtained via bifurcation of a vortex state,
in which the particles are localized in a single site.
See \Fig{fg00}e for pedagogical illustration. 
%
Other notable fixed-points correspond to single-depleted-well states 
in which one site is empty, while the remaining two sites are equally occupied by the particles.
For ${M>3}$ rings there are off-axis fixed points 
that support spatially modulated vortex states \cite{Pethick}.

\section*{The $M{=}3$ ring}

We focus our attention on the central region of energies, where the middle vortex state ($m=1$) is located.
Examples for phase-space trajectories at this energy are displayed in \Fig{fg2}, 
and will be further discussed below. First we examine the linear stability of the pertinent vortex fixed-point.

\sect{BdG analysis} 
The characteristic equation ${\det(\lambda - \mathbb{J} \mathcal{A})=0}$ 
has two trivial eigenvalues $\lambda=0$ that reflect the 
the constant of motion $\mathcal{N}$, and therefore excluded. 
The other four eigenvalues are the solution of ${\lambda^4 +b\lambda^2+c =0}$ 
with
\beq  
c &=& \sin^2 \left(\frac{\Phi}{3}-\frac{\pi}{6}\right) \left[ u- \frac{3-12 \sin^2 \left(\frac{\Phi}{3}-\frac{\pi}{6}\right)}{4\sin \left(\frac{\Phi}{3}-\frac{\pi}{6}\right)} \right]^2 
\\
b &=& \frac{3}{2} +3 \sin^2 \left(\frac{\Phi}{3}-\frac{\pi}{6}\right) +2 u \sin \left(\frac{\Phi}{3}-\frac{\pi}{6}\right) 
\eeq
A fixed-point is energetically (meta) stable if it 
seats at a (local) minimum or maximum of $H(z)$. 
At the energetic-stability border, 
where the fixed-point becomes a saddle, 
we have ${\det(\mathcal{A})=\det(\mathbb{J}\mathcal{A})=0}$, 
hence the border is determined by ${c=0}$, 
leading to the energetic-stability condition
\beq
u \ \ > \ \ \frac{3-12 \sin ^2 \left(\frac{\Phi}{3}-\frac{\pi}{6}\right)}{4\sin \left(\frac{\Phi}{3}-\frac{\pi}{6}\right)} 
\eeq
The fixed-point becomes dynamically unstable if the eigenvalues acquire 
a real part, which is the so-called Lyapunov exponent. 
This happen when ${b^2-4 c<0}$, leading to dynamical instability in the region 
\beq
u > \frac{9}{4} \sin \left(\frac{\pi}{6}-\frac{\Phi}{3}\right) \ \ \ \ \text{with} \ \ \ \ \Phi < \frac{\pi}{2}
\eeq 
In principle for ${b^2-4c>0}$ the condition $b<0$ would 
imply an additional dynamical instability regime.  
But here ${b<0}$ occurs inside the region of ${b^2-4c<0}$. 
The stability borders are demonstrated in \Fig{fg1}a.

\begin{figure}
\begin{center}

\includegraphics[width=8cm]{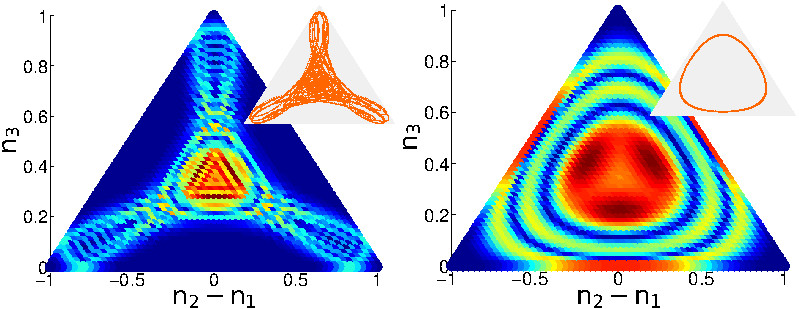}



\caption{ \label{fg3} 
{\bf Wavefunctions of exotic vortex states.} 
Chaotic vortex state (left); and breathing vortex state (right) of an $N{=}60$ trimer are imaged:  
the probability ${ |\psi(r)|^2 }$, see text, is color-coded blue (zero) to red (max).
The underlying classical dynamics (insets) is chaotic or periodic respectively.
The parameters ${(\Phi,u)}$ are ${(0.36 \pi, 1.3)}$ and ${(0.05 \pi, 1.5)}$. 
Note that a regular vortex (not displayed) would be represented 
by a simple hump at the central ($n_1{=}n_2{=}n_3$) fixed-point. 
The occupation axis is scaled (${n\mapsto n/N}$). 
} 
\end{center}
\end{figure}

\sect{Beyond BdG} 
In \Fig{fg1}a we observe macroscopically large currents beyond the expected region. 
This has been noted in \cite{sfs} without explanation.  
In particular we see that supefluidity survives in the limit $\Phi\rightarrow0$, 
contrary to the expectation \cite{Ghosh,paraoanu} that is based on the traditional stability argument. 
We can plot the wavefunction ${|\Psi(r)|^2 = \left|\langle r | E_{\alpha} \rangle\right|^2}$ 
of an eigenstate that supports superfluidity. Here the components of~$r$
are the coordinates ${r_{\parallel}=(n_2-n_1)/N}$ and ${r_{\perp}=n_3/N}$.
The wavefunction of a standard {\em regular} vortex is merely a hump at the 
symmetry point ${n_1=n_2=n_3=N/3}$. In \Fig{fg3} we 
display examples for the wavefunctions of non-standard vortex states: 
a {\em chaotic} vortex state, and a {\em breathing} vortex state.
The terms "chaotic" and "breathing" is related 
to the underlying classical dynamics [options (ii) and (iii) below].
 
\rmrk{Launching trajectories at the vicinity of the vortex fixed-point 
it has been demonstrated that both stable and unstable oscillations can emerge \cite{trimer6}.}
Specifically we encounter the following possibilities:
{\bf (i)}~the trajectories are locked in the vicinity of the vortex fixed point;
{\bf (ii)}~the trajectories are quasi-periodic in phase-space;
{\bf (iii)}~the trajectories are chaotic but uni-directional. 
The above possibilities are pedagogically illustrated by the cartoon of \Fig{fg00}.
Poincare sections of the trajectories (see below) reveal that 
a regular vortex state is supported by a regular island around 
the fixed point (case~i); a breathing vortex is supported by 
a secondary island that has been created via bifurcation (case~ii);
while a chaotic vortex state is supported by a 'chaotic pond' 
of clockwise motion that does not mix with the anti-clockwise motion (case~iii). 
Consequently the motion may become chaotic, but stay uni-directional,  
and superfluidity persists contrary to the common expectation.

\sect{The route to chaos} 
As observed in \Fig{fg1}a superfluidity is quite robust. 
The current diminishes only in the vicinity of what we call ``swap transition", 
which is indicated in the figure by dotted line. In the Methods section  
we derive an explicit expression for the transition line:
\be{12} 
u \ = \  18\sin\left(\frac{\pi}{6}-\frac{\Phi}{3}\right) 
\eeq 
At the transition the two separatrixes that dominate 
the structure of phase-space coalesce, and consequently 
a global chaotic-sea is formed.
The details of this transition are provided below.
Such type of separatrix overlap bifurcation has been once encountered 
in molecular physics studies \cite{Svitak}, but not in chaotic context.

\sect{Phase-space tomography} 
In \Fig{fg2} we plot the spectrum of the trimer Hamiltonian for 
representative values of ${(\Phi,u)}$ that are indicted in \Fig{fg1}a. 
We also plot in each case the Poincare sections at the energy of the $m=1$ vortex fixed-point 
(see Methods section for details). 
From the quantum spectrum we can easily deduce 
the phase-space structure at {\em any other energy}.
One can call it ``quantum phase-space tomography".
Consider for example \Fig{fg2}c. We can easily correlate 
the largest current states with the red (upper) island; the secondary
group of large current states with the yellow (left) island; 
and the small current states with the green chaotic sea.
Additional information can be extracted from the {\em purity} of the states. 
For precise definition see \cite{sfs}.  
Here it is enough to say that ${S=1}$ means that all the particle 
are condensed in a single orbital, while ${S<1}$ means that  
the state is fragmented into $1/S$ orbitals. For ergodic state $1/S\sim M$. 
Points in the spectrum are color-coded from black ($S\sim 1$) to purple ($S \sim 1/M$).   
One observes that the non-standard vortex states have high but not perfect purity.

By inspection of \Fig{fg2} we observe the following regimes in the diagram of \Fig{fg1}a: 
Regime (S) stands for simple phase-space structure with energetically-stable 
clockwise (``red") and anti-clockwise (``blue") islands 
that are separated by a forbidden region.
In regime (B) we have two regular regions of clockwise motion, 
and ``blue separatrix" that supports anti-clockwise motion. 
As we go up in $u$ the blue separatrix becomes a chaotic sea.
In regime (D) the middle vortex bifurcates, 
while the other clockwise island remains regular.
In regime (D') the ``vortex separatrix'' swaps with the ``blue separatrix". 
This swap is clearly demonstrated 
as we go from \Fig{fg2}d to \Fig{fg2}e.
The border between regimes (D) and (D') is shown as a dotted 
line in \Fig{fg1}a, see \Eq{e12}. 
Along this line the two separatrices coalesce. 
Crossing to regime (B') the bifurcation that is responsible 
to the ``blue separatrix" is undone, and eventually 
we can go back to the (S) regime via a simply-connected (A) regime 
that has a simple structure with no separatrix.   
     
In region (B) the vortex is not energetically stable:  
it is located on a saddle point in phase-space. 
Nevertheless dynamical stability is maintained. 
In region (D) the vortex is no longer dynamically stable, 
and the trajectories at the vicinity of the vortex are chaotic.
Still the motion is confined by KAM tori 
within a ``chaotic pond", and therefore remains uni-directional. 
In the vicinity of the swap,  as we go up in~$u$, 
the chaotic pond becomes a chaotic sea, 
and the superfluid current is diminished.

Upon quantization the chaotic pond 
can support a ``chaotic vortex state", 
which has been illustrated in \Fig{fg3}a.  
A second class of large current states 
are supported by stable {\em periodic-orbits} (POs)  
that has been bifurcated from the stationary 
vortex fixed-point, once the latter lost stability.  
These POs are elliptic fixed-points 
of the Poincare section, see  \Fig{fg2}d. 
Upon quantization the associated islands 
can support a ``breathing vortex state", see \Fig{fg3}b.

\section*{The $M{>}3$ rings}

Considering a no-rotating device, the traditional stability 
argument \cite{Ghosh,paraoanu} implies marginally stable superfluidity 
for an $M{=}4$ device, and stability if~$u$ is large enough for an $M{>}4$ device.
These observations are implied by the stability borders 
that are plotted in \Fig{fg1}bc. The explicit expressions 
for the stability borders are provided in the following paragraph. 

\sect{BdG analysis} 
In the Methods section we derive an explicit expression \Eq{e170}
for the Bogoliubov frequencies $\omega_{q,\pm}$ of the $k_m$ vortex.
The frequencies are indexed by $q=(2\pi/M)\times\text{integer}$.     
As we move in the $(\Phi,u)$ regime-diagram 
the $\omega_{q,\pm}$ go through zero as we cross the following lines
\beq
u = \frac{M}{\cos \left(k_m-\frac{\Phi}{M} \right) }
\left[ \cos^2 \left(\frac{q}{2}\right) - \cos^2 \left(k_m-\frac{\Phi}{M} \right)  \right] 
\eeq
The border of the stable regime is determined by the first line that is encountered, 
which corresponds to the minimal value ${q=2\pi/M}$. 
The dynamical stability is lost once one of  
the frequencies become complex. This happens when 
either of the following two expressions equal zero
\beq
\left\{ 
\cos\left( k_m {-} \frac{\Phi}{M}\right), \ \   
\frac{u}{M} + \cos \left( k_m {-} \frac{\Phi}{M} \right) \sin^2 \left( \frac{q}{2} \right)
\right\} \ \ \ \ \ 
\eeq
Again we have to substitute ${q=2\pi/M}$, which is associated with the first encounter.
The explicit results for $M=3$ are easily recovered.

\sect{Beyond BdG} 
Looking on the numerically determined current (\Fig{fg1}b) one observes 
that superfluidity can persist slightly beyond the dynamical stability border. 
But much more conspicuous is the diminishing of superfluidity within a large 
region where the BdG analysis predicts dynamical stability. 
We find (see below) that the latter effect is related to Arnold diffusion. 
Namely, if $d{>}2$, the $d$~dimensional KAM tori in phase space 
are not effective in blocking the transport on the ${2d{-}1}$
energy shell (as discussed in the Introduction). 
As~$u$ becomes larger this non-linear leakage effect is enhanced, 
stability of the motion is deteriorated, and the current is diminished.

\begin{figure}
\begin{center}

\includegraphics[width=8cm]{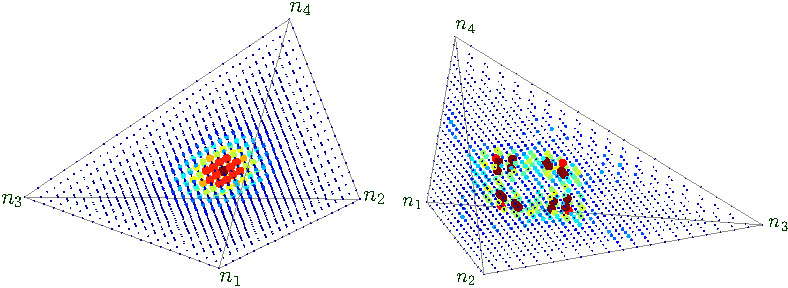}

\caption{ \label{fg3a} 
{\bf Wavefunctions of vortex state of $M{>}3$ circuit.}
Displayed are regular vortex state (left) and irregular vortex state (right) 
of $M{=}4$ ring with $N{=}20$ particles.
The axes are as in \Fig{fg3} with one extra dimension, and the color code is the same. 
The parameters ${(\Phi,u)}$ are ${(0.2 \pi, 0)}$ and ${(0.2 \pi, 3)}$ respectively. 
Note that a regular vortex is represented 
by a simple hump at the central ($n_1{=}n_2{=}n_3{=}n_4$) fixed-point, 
whereas an irregular vortex has a richer structure that 
reflects the fragmented phase-space structure.  
} 

\end{center}
\end{figure}

At this point it is helpful to distinguish between
{\em strict} dynamical stability and {\em linear} dynamical stability~\cite{LL}.
For $M{>}3$ the latter does not imply the former (all regions of phase 
space are interconnected via ``Arnold diffusion" irrespective of their linear stability). 
As we go up in $u$ the chaos becomes predominant, 
and consequently energetic-stability rather than dynamical-stability 
becomes the relevant criterion. 
It follows from this distinction that for $M{>}3$ one has to
distinguish between regular and irregular vortex states. 
This distinction is demonstrated in \Fig{fg3a}.  
A regular vortex is represented by a simple hump 
at the central ($n_i=N/M$) fixed-point, 
whereas an irregular vortex has a richer structure 
that reflects the underlying fragmented phase-space structure.

In order to verify the above semiclassical reasoning, 
we try in \Fig{fg4} to reconstruct the quantum regime-diagram 
via classical simulations.  This reconstruction
provides a qualitative proof for the semiclassical reasoning, 
and furthermore demonstrates the $N$ dependence 
of the the superfluity regime diagram. 
Namely, we launch a Gaussian cloud of trajectories that
have an uncertainty width that corresponds to $N$. 
The fraction of trajectories that escape is used 
as a measure for the stability. The practical criterion 
for escape is having the average current $I(t)$ getting below 
some threshold $I_{\infty}$ within some time $t_{\infty}$.
In principle $t_{\infty}$ should be the Heisenberg time (proportional to~$N^d$), 
and $I_{\infty}$ can be (say) half $I_m$. In practice the result 
is not sensitive to these thresholds.  
Regions where the current diminishes in-spite of BdG dynamical stability, 
and does not recover even if~$N$ is increased (smaller uncertainty width), 
establish the relevance of Arnold diffusion.

\section*{Discussion} 

\rmrk{A recent experiment with a toroidal ring \cite{n5} has demonstrated hysteresis 
in a quantized superfluid ‘atomtronic’ circuit. The procedure was 
to prepare a stable vortex state at rotation frequency that corresponds to~$\Phi_1$, 
and then to check its stability after changing to~$\Phi_2$. 
The theory there could be regarded as an extension of the 
traditional ``energetic-stability" reasoning that underlays the Landau criterion. 
Our work has been motivated by the following question: {\em what would be the results 
of the same type of experiment, if the toroidal ring had several weak links, or if 
it were replaced by a discrete $M$~site ring}. The realization of such ring 
that is described by \Eq{e1} has been discussed in \cite{ref9,e16,e17}. 
We argue that the chaotic nature of such circuit requires to go 
beyond the traditional analysis; else one would not be able 
to predict the borders of the stability-regime in the hysteresis loop.}
On the theoretical level we have highlighted a novel type of superfluidity that is 
supported by irregular or chaotic or breathing vortex states. 
Such states are supported by fragmented regions in phase-space ($M{>}3$),  
or by chaotic ponds ($M{=}3$), or by periodic-orbits respectively, 
hence they are missed by the traditional BdG analysis. Furthermore 
we have highlighted the limitations of the {\em linear} stability 
analysis for high dimensional chaos ($M{>}3$).

As a secondary message, we would like to emphasize 
that the gross features of the classical phase-space 
can be easily extracted from the spectrum of the quantized Hamiltonian. 
To get the same information via classical analysis would be an extremely 
heavy task that would require generation of many trajectories 
in numerous phase space regions, 
on each possible energy shell, 
as opposed to our ``quantum phase space tomography" 
which requires a single diagonalization of a finite matrix. 
If Nature were classical, Quantum Mechanics still would 
be invented as a valuable tool, just for the purpose of analysing 
mixed complex dynamics.


\section*{Methods}

\sect{Poincare sections}
Starting with the Hamiltonian \Eq{e1} with $M=3$, written in terms of action-angle variables, 
the classical dynamics is generated by the equation
\beq   
i\frac{\partial a_i}{\partial t} \ \ = \ \ 
u |a_i|^2 a_i - \frac{1}{2} \left[  e^{i \Phi/3} a_{i-1}   + e^{-i \Phi/3} a_{i+1} \right]
\eeq
with scaled units such that ${K=N=1}$. 
We solve this equation numerically. 
For plotting of trajectories it is convenient to use 
the coordinates $(n_1-n_3,\varphi_1-\varphi_3)$ and $(n_3-n_2,\varphi_3-\varphi_2)$.
The section chosen is ${n_3-n_2=0}$, at the energy of the ${m=1}$ vortex ${ E = (u/6) - \cos[ (2\pi/3){-}(\Phi/3) ] }$.
Given a phase space section point $(n_1-n_3,\varphi_1-\varphi_3)$, 
the equation ${H(z) = E}$ has either zero or two solutions 
for the remaining coordinate ${\varphi_3-\varphi_2}$. 
This implies that the Poincare section has two sheets.
For presentation purpose we pick the sheet where  
velocity ${\partial_t (n_3-n_2)}$ has a larger value.

\begin{figure}
\begin{center}

\includegraphics[width=1\hsize]{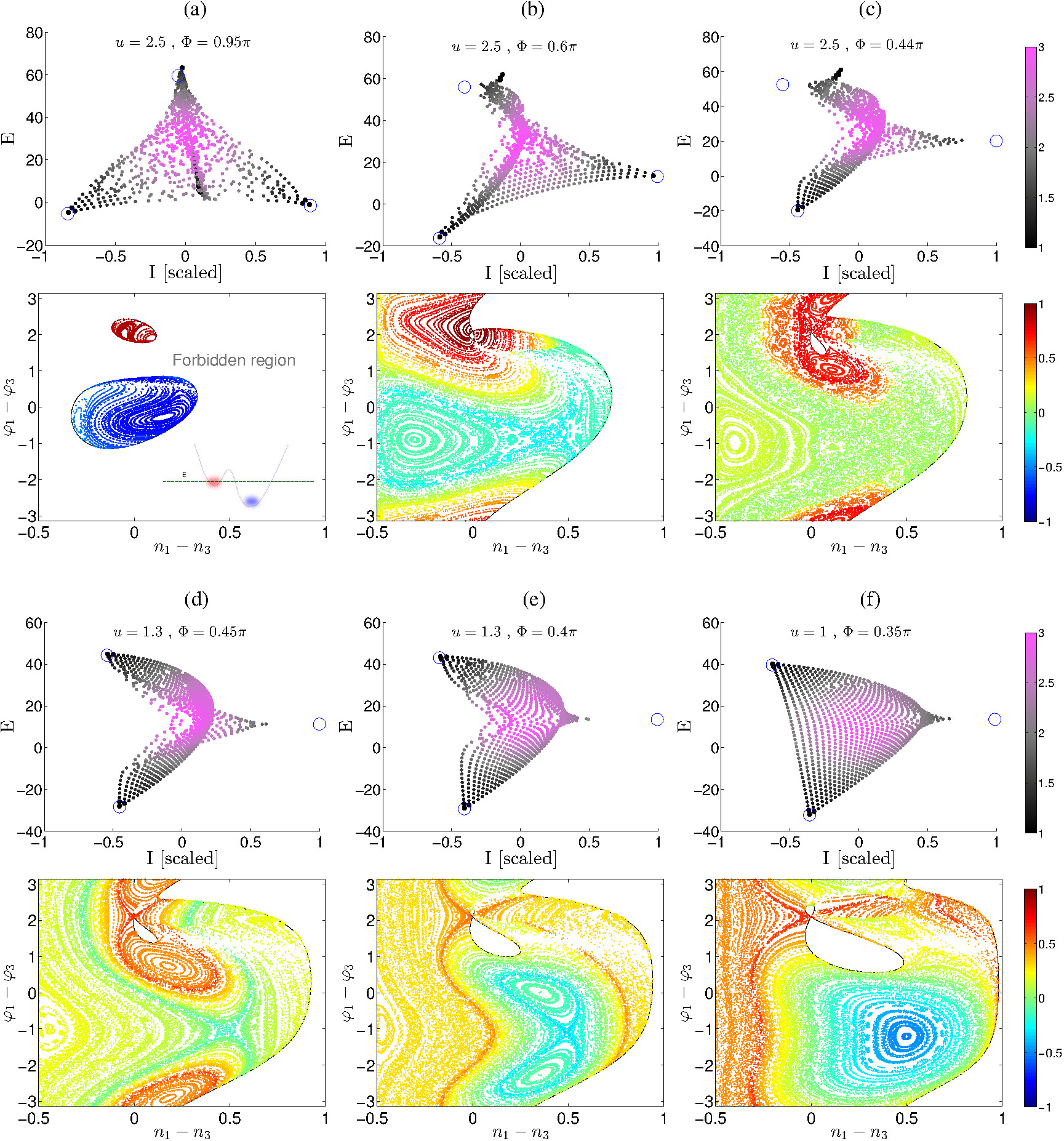}

%
%

\caption{\label{fg2}
{\bf Quantum spectrum and phase-space landscape.}  
Representative quantum spectra of an $N=42$ trimer 
for representative $(\Phi,u)$ values. 
Panels (a-f) are for 
$(0.95 \pi,2.5)$; $(0.6 \pi,2.5)$; $(0.44 \pi,2.5)$; 
$(0.45 \pi,1.3)$; $(0.4 \pi,1.3)$; $(0.35 \pi,1)$.
Each point represents an eigenstate color-coded by 
its purity (black $(1/S)\sim1$ to purple $(1/S)\sim3$), 
and positioned according to its energy $E_{\alpha}$ 
and its scaled current $I_{\alpha}/(NK/M)$.
In each case an $n_3{-}n_2{=}0$ Poincare section 
at the energy of the $m=1$ vortex is displayed 
(with the exception of (a) where it is for a slightly 
shifted energy, else the red island would shrink into a point). 
The vortex location is ${(n_1{-}n_3{=}0, \ \varphi_1{-}\varphi_3{=}2\pi/3)}$.
The solid black line marks the borders of the 
allowed phase-space regions.  
The color code represents the averaged current 
for each classical trajectory.} 

\end{center}
\end{figure}

In \Fig{fg2} we plot Poincare sections at the energy of the $m=1$ vortex fixed-point.
The section is $n_3-n_2=0$. For convenience we use the canonical coordinates $(n_1-n_3,\varphi_1-\varphi_3)$ 
and a scaled particle number ${n\mapsto n/N}$.
Note that the $m=1$ vortex fixed-point is always located at $(0,2\pi/3)$. 
The boundary of the allowed region is marked by a black line. 
Each trajectory in the Poincare section is a set of points 
that share the same color. The color reflects the average current, 
which is calculated by taking a time average over~$\mathcal{I}$.

We have here a $d=2$ system, so the energy shell is 3D, 
while the Poincare section is 2D. 
According to KAM, some regular motion survives: these are the KAM tori. 
In the PS they look line lines that divide  
the motion into ``territories". For example 
in \Fig{fg2}b we see that most of the motion is regular. 
We have there some bounded chaotic motion in the vicinity 
of the blue separatrix, but near the vortex fixed-point 
the motion is predominantly regular (dense set of KAM tori).    
In \Fig{fg2}c most of the KAM tori have been destroyed.
Still the remaining KAM tori separate the red chaotic pond 
from the green chaotic sea.
If these tori were destroyed, the motion would 
become globally chaotic (which is not the case).
\\

\sect{Stability analysis for general $M$} 
It is convenient to do the general stability analysis using the common many-body ``quantum" notations.
The canonical variables are $a_j$ and its conjugate instead of $(\bm{\varphi},\bm{n})$. 
Then we transform to $b_k$ that are defined via the substitution  
\beq
a_j \ \ = \ \ \frac{1}{\sqrt{M}} \sum_j \eexp{ikx_j} \ b_k
\eeq
The BHH takes the form
\beq
\mathcal{H} \ \ = \ \ \sum_{k} \epsilon_k b_k^{\dag} b_k \ \ +  \ \ \frac{U}{2M} \sum_{k_1+k_2=k_3+k_4}  b_{k_4}^{\dag}b_{k_3}^{\dag}b_{k_2}b_{k_1}
\eeq
where $\epsilon_k = -K \cos (k-\Phi/M)$. 
The fixed-point that is associated with ${k_m=(2\pi/M)m}$ 
is $b_k^{(0)}=\sqrt{N} \ \delta_{k,k_m}$.   
The linearized Hamiltonian around this fixed-point is separable:
\beq
\mathcal{H} \ \ \approx \ \ \sum_{k (\ne k_m)} \varepsilon_k b_k^{\dag} b_k \ \ + \ \ \Delta \sum_{q>0} \left( b_{k_m+q}^{\dag}b_{k_m-q}^{\dag} + b_{k_m+q}b_{k_m-q}  \right) \ \ \equiv \ \ \sum_q \mathcal{H}^{(q)}
\eeq 
where $q=(2\pi/m)\times\text{integer}$ are the possible differences in $k$ values, and   
\beq
\varepsilon_k = \epsilon_k - \epsilon_{k_m} + \frac{NU}{M}, \ \ \ \ \ \ \ \ \Delta=\frac{NU}{M}
\eeq
and
\beq
\mathcal{H}^{(q)} \ \ = \ \ \left(\varepsilon_{k_m+q} b_{k_m+q}^{\dag} b_{k_m+q} + \varepsilon_{k_m-q} b_{k_m-q}^{\dag} b_{k_m-q}\right) + \ \  \Delta \ \left(b_{k_m+q}^{\dag} b_{k_m-q}^{\dag} +b_{k_m+q} b_{k_m-q}\right) 
\eeq 
After Bogoliubov transformation it takes the form 
\beq
\mathcal{H}^{(q)} \ \ = \ \ \omega_{q,+} c_{q,+}^{\dag} c_{q,+} + \omega_{q,-} c_{q,-}^{\dag} c_{q,-}
\eeq 
where the Bogoliubov excitation frequencies are 
\beq
\omega_{q,\pm} \ \ = \ \ 
\left[ \left(\frac{\varepsilon_{k_m+q}+\varepsilon_{k_m-q}}{2} \right)^2 - \Delta^2 \right]^{1/2}
\ \ \pm \ \ \left(\frac{\varepsilon_{k_m+q}-\varepsilon_{k_m-q}}{2} \right) 
\eeq
Note that the familiar textbook expression for the phonon excitation of the $\Phi=0$ ground state 
is obtained upon the substitution ${k_m=0}$. For the problem under study we get
\be{170}
\omega_{q,\pm} \ = \  2 K \sin \left(\frac{q}{2} \right)
\left[ 
\frac{u}{M} \cos \left(k_m- \frac{\Phi}{M}\right)
+\cos^2 \left(k_m-\frac{\Phi}{M}\right) \sin^2 \left(\frac{q}{2}\right)\right]^{1/2} 
\ \pm \ \ K \sin \left(q \right) \sin \left(k_m-\frac{\Phi}{M} \right) 
\eeq
where $u=NU/K$. 
\\

\begin{figure}[h!]
\begin{center}

\includegraphics[width=10cm]{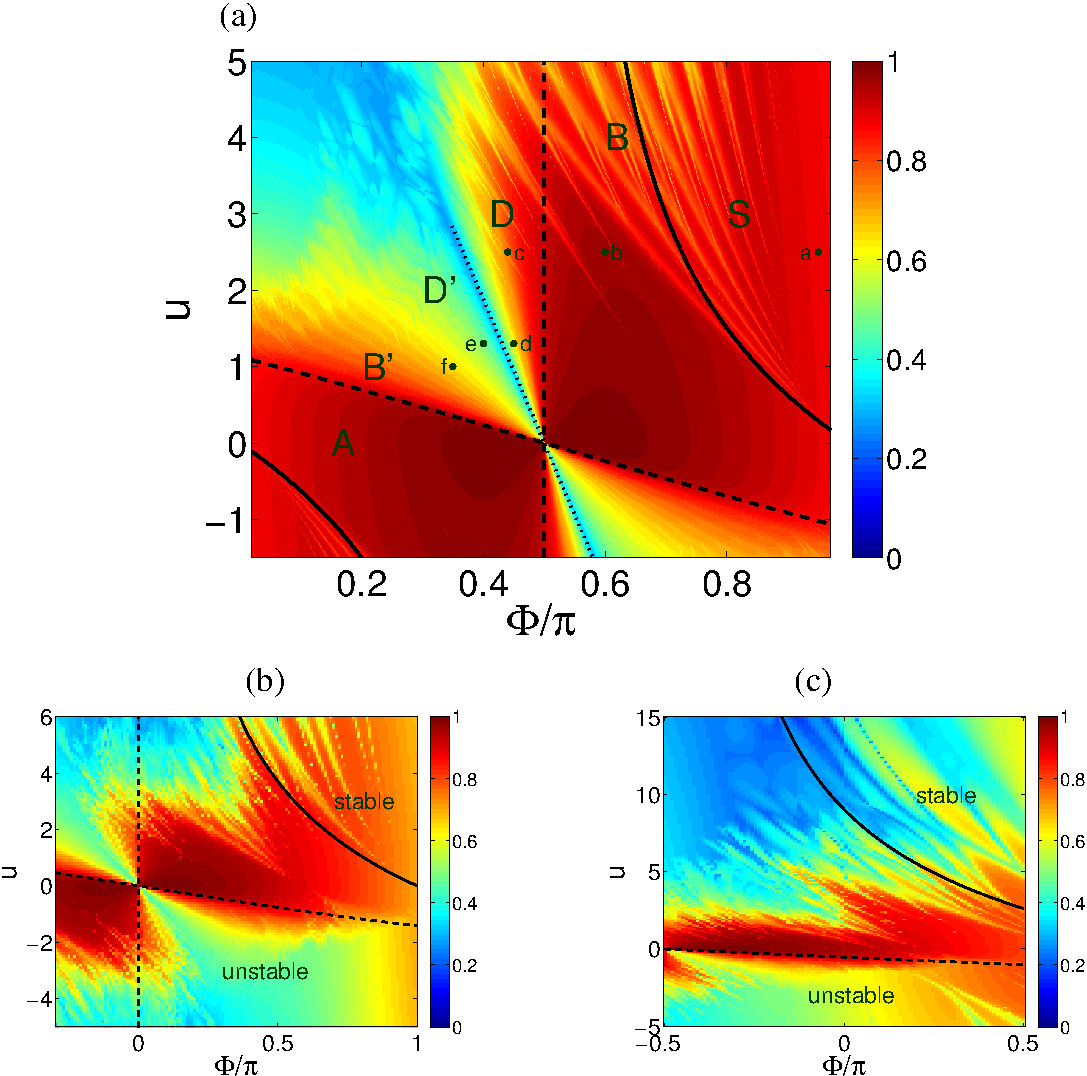}

%
%
%

\caption{ \label{fg1} 
{\bf Superfluidity regime diagram for $M$ site circuits. }
(a)~$M{=}3$ ring with ${N{=}37}$ particles;     
(b)~${M{=}4}$ ring with ${N{=}16}$ particles;  
(c)~${M{=}5}$ ring with ${N{=}11}$ particles.
The model parameters are $(\Phi,u)$. 
The~$I$ of the state that carries maximal current 
is imaged at the background. 
We observe that the stability regions (large current) 
are not as expected from the linear stability analysis: 
the solid line indicates the energetic-stability border;  
the dashed lines indicate the dynamical stability borders. 
For clarity we also include a negative $u$ region 
which is in fact a duplication of the upper sheet. 
In~(a) the dotted line indicates the ``swap" transition (see text);  
and the dots labeled (a-f) mark $(\Phi,u)$ coordinates 
that are used in \Fig{fg2} to demonstrate the different regimes.
} 

\end{center}
\end{figure}

\begin{figure}
\begin{center}

\includegraphics[width=8cm]{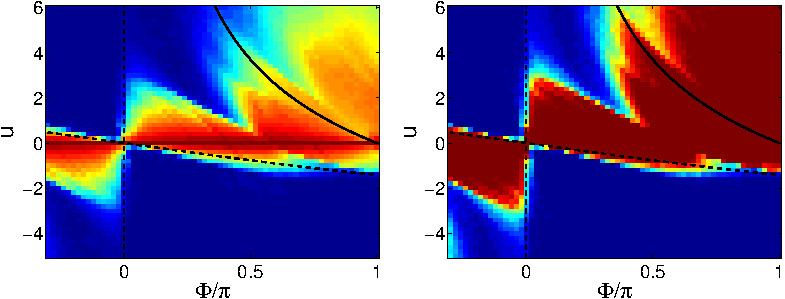}

\caption{ \label{fg4} 
{\bf Semiclassical reproduction of the regime-diagram.} 
The calculation is for $M{=}4$ circuit, 
and should be compared with \Fig{fg1}b.
The lines are the BdG stability borders.
Given ${(\Phi,u)}$ we launch a Gaussian cloud of trajectories that
have an uncertainty width that corresponds to~$N$. 
The fraction of trajectories (blue$\sim100\%$ to red$\sim0\%$) 
that escape is used as a measure 
for stability (see text for details).  
Results are displayed for clouds that have 
uncertainty width ${\Delta\varphi \sim \pi/2}$ (left) 
and ${\Delta\varphi \sim \pi/4}$ (right).} 

\end{center}
\label{f3}
\end{figure}

\sect{Swap scenario}
Inspecting \Fig{fg1} one observes that superfluidity diminishes 
in the vicinity of the indicated dotted line. 
It turns out that the swap scenario is originated from a global 
non-linear resonance. This can be established by inspection 
of Poincare sections (see below): at the transition two separatrixes swap in phase-space.
In order to derive the resonance condition we rewrite the Hamiltonian \Eq{e1} using:
\beq   
b_0 = \frac{1}{\sqrt{3}} \left(a_1 +a_2+a_3\right), \ \ \ \ \ \ \ \ 
b_{\pm} = \frac{1}{\sqrt{3}} \left(a_1 e^{\pm i\frac{2\pi}{3}} +a_2 e^{\pm i\frac{4\pi}{3}}+a_3\right)  
\eeq
Defining $\omega_0 = -K\cos (\Phi/3)$, and $\omega_{\pm} = -K\cos (\pm 2\pi/3 - \Phi/3)$, this leads to:
\beq \nonumber
\mathcal{H} &=& \omega_0 n_0 +  \omega_+ n_+ + \omega_- n_- 
\ \ + \ \ \frac{U}{6} \left[ n_0^2+n_+^2+n_-^2 +  4(n_0 n_+ + n_0 n_- + n_+ n_-) \right]   
\\ 
&+& \frac{U}{12} \left[ 
     ( b_+ ^{\dag}b_+ ^{\dag}b_- b_0 + b_0 ^{\dag} b_- ^{\dag}b_+ b_+)
   + ( b_0 ^{\dag}b_0 ^{\dag}b_+ b_- + b_- ^{\dag} b_+ ^{\dag}b_0 b_0)
   + ( b_- ^{\dag}b_- ^{\dag}b_0 b_+ + b_+ ^{\dag} b_0 ^{\dag}b_- b_-)   
  \right]
\eeq
We consider the subspace of states with ${n_{0} - n_{-} =0 }$, 
and keep only the resonant coupling $(b_{+}^{\dag})^2 b_{-} b_0$ and its conjugate.
Then we define the reaction coordinate is $J_z = (1/4)(2n_{+}-N)$, 
associated ladder operators $J_{\pm}$, and hooping generator $J_x=(J_{+}+J_{-})/2$,  
such that the Hamiltonian takes the form:
\beq
\mathcal{H} \approx \left( 2\omega_{+} - \omega_{-} - \omega_{0} - \frac{UN}{6} \right) J_z
\ - \ U J_z^2 \ + \ \frac{U}{3} \left[ (N/4)^2 - J_z^2  \right]^{1/2} \ J_x \ + \ \const
\eeq
The non-linear resonance happens if the first term vanishes, leading to \Eq{e12}.

\clearpage

\vspace*{-5mm}

\hidea{

\section*{Acknowledgements}

This research has been supported by  by the Israel Science Foundation (grant Nos. 346/11 and 29/11) 
and by the United States-Israel Binational Science Foundation (BSF).

\section*{Author contributions statement}

All authors have contributed to this article. 
The numerical analysis and the figures have been prepared by G.A., 
while the text of Ms has been discussed, written and iterated jointly by the 3 authors.

\section*{Additional information} 

The authors declare that they have no competing financial interests.

}

\clearpage

\begin{thebibliography}{99}
\bibitem[a]{rmA}
While quantum mechanics allows tunnelling
through classically-forbidden regions into the allowed quasi-continuum, 
this 'phase slip' process \cite{e9,Altman,e13,e15} is much slower  
and is therefore ignored in the semi-classical analysis.
\end{thebibliography}
\end{document}